\pdfoutput=1
\documentclass[aps,reprint,nofootinbib,superscriptaddress,onecolumn,notitlepage]{revtex4-1}
\usepackage{custom}
\begin{document}

\title{Heavy-dense QCD, sign optimization and Lefschetz thimbles}

\author{G\"{o}k\c{c}e Ba\c{s}ar}
\email{gbasar@unc.edu}
\author{Joseph Marincel}
\email{marincel@unc.edu}
\affiliation{Department of Physics and Astronomy, University of North Carolina, Chapel Hill, North Carolina 27599, USA}

\date{\today}

\begin{abstract}   
We study the heavy-dense limit of QCD on the lattice with heavy quarks at high density. The effective three dimensional theory has a sign problem which is alleviated by sign optimization where the path integration domain is deformed in complex space in a way that minimizes the phase oscillations.  We simulate the theory via a Hybrid-Monte-Carlo, for different volumes, both to leading order and next-to-next-to leading order in the hopping expansion, and show that sign optimization successfully mitigates the sign problem at large enough volumes where usual re-weighting methods fail. Finally we show that there is a significant overlap between the complex manifold generated by sign optimization and the Lefschetz thimbles associated with the theory.  

\end{abstract}

\maketitle

\section{Introduction}
\label{sec:intro}

Mapping the phase diagram of Quantum Chromo Dynamics (QCD), the theory that governs the strong nuclear force, has been an outstanding problem for many decades. The main challenge in a nutshell is that such a task requires non-perturbative methods which almost always involve a numerical stochastic, ``Monte-Carlo" component where the path integral is sampled statistically. Even though such non-perturbative methods have been very successful in computing the thermodynamic properties of QCD at small densities \cite{HotQCD:2014kol,BORSANYI201499},
for most theories at finite density, including QCD, their applicability is severely limited by the ``sign-problem" which arises when path integral measure is not positive definite and therefore cannot be interpreted as a probability measure.  In many cases such as QCD at finite density, the underlying action is complex and the Boltzmann weight, $e^{-S}$, leads to severe phase oscillations in the path integral which become exponentially rapid in the large volume / low temperature limit \cite{deForcrandreview,Philipsen:2007aa,KARSCH200014,Aarts_2016}. These oscillations make the numerical computation of the path integral practically impossible.    

Over the years, there has been many attempts to tackle the sign problem. One set of ideas stem from complexification of fields where instead of sampling the configurations on the original domain of the path integral (for example $SU(N_C)$ for QCD), one samples on a complexified domain (a subset of $SL(N_C)$ for QCD) where the phase oscillations are milder and therefore can be dealt with using conventional methods.  Of course, the deformation to the complexified fields must not change the value of the path integral. An example of a complex domain over which the path integral has milder phase oscillations is the Lefschetz thimble decomposition, which is a multi-dimensional generalization of expressing a one-dimensional integral as a linear combination of steepest descent contours \cite{pham,Lefschetz:1975ta}. Just like steepest descent contours, the phase over each Lefschetz thimble is stationary, rendering the phase oscillations milder \cite{Cristoforetti:2012su,Cristoforetti:2012uv,Fujii:2013sra}. However finding the correct linear combination of thimbles for a given set of parameters in the theory and sampling is generally challenging. To overcome these difficulties various methods have been proposed where instead of the thimbles, one samples over other complex domains which still have milder phase oscillations than the original domain but easier to construct and sample compared to thimbles \cite{Alexandru:2020wrj}.  
 
A more direct approach to mitigating the sign problem by complexification is to convert it into an optimization the problem where one minimizes the sign oscillations within a family of complex domains parameterized by a set of auxiliary variables. In a nutshell, an observable, such as the average phase, that measures the strength of the sign oscillations is minimized within that set of parameters that describe the complexified domain. The set of ideas that stem from this approach goes under the name ``sign optimization" and has been applied to various QCD-like models \cite{Mori:2017pne, Alexandru:2018fqp,Kashiwa:2019lkv}.  
More broadly, this idea has also been used to mitigate signal-to-noise problems in gauge theories even in the absence of a sign problem \cite{Detmold:2021ulb}

In this paper, we implement sign optimization in the heavy-dense limit of QCD which is captured by an effective theory whose degrees of freedom describe heavy quarks at very high density. Even though this is a somewhat academic limit of QCD, it does inherit the sign problem from QCD which becomes exponentially severe with the volume/inverse temperature. Furthermore the degrees of freedom are elements of $SU(N_C)$ the complexification of which can be generalized to QCD. Furthermore, the simplified mathematical structure of the theory in this limit allows analytical expressions which can be directly compared with the Monte-Carlo results obtained from sign optimized lattice simulations. Finally, because of the same reason, the simulations are not as demanding as QCD in terms of computational resources. All these factors combined make heavy dense QCD an ideal test ground for implementing sign optimization with the goal of applying it to QCD in the future. Along similar lines to this work, it has been studied via Lefschetz thimbles \cite{Zambello:2018ibq} and Complex Langevin \cite{Attanasio:2015Cg,Aarts:2016qrv} in the context of the sign problem. 

In Ref. \cite{Basar:2022cef} we studied the one-dimensional limit of QCD in the context of sign optimization. This paper is a natural continuation of the work along those lines. One crucial difference, however, is that heavy-dense QCD is a three-dimensional effective theory where the sign problem grows exponentially with the volume, allowing us to analyze the performance of sign optimization as a function of volume; which was not possible in one-dimensional QCD.  At the same time, as we discuss further in the paper, the existence of \textit{complex} saddle points and the relations between the Lefschetz thimbles attached to these saddles and the sign optimized manifolds is similar to what we have observed in one-dimensional QCD.

The rest of the paper is organized as follows. In Sections \ref{sec:hd} and \ref{sec:sign_optimization} we briefly recapitulate the heavy-dense effective theory of QCD, and sign optimization respectively. Our results are presented in Section \ref{sec:results}. We then compare the complexified domains obtained by sign optimization to Lefschetz Thimbles in Section \ref{sec:thimbles}. Finally we discuss our results and present our conclusions in Section \ref{sec:conclusions}.

%


\section{Heavy Dense QCD}
\label{sec:hd}

We consider the cold and dense limit of QCD with heavy quarks \cite{Fromm:2011qi,Fromm:2012eb,Langelage:2014vpa}. In this limit, the quarks cannot move due their large mass, but do not decouple entirely either, due to their high density. The resulting effective theory is three dimensional whose degrees of freedom are Polyakov loops, $P(\vx)$, that describe these stationary quarks. More precisely, to leading order in this heavy-dense limit, the Dirac determinant (for Wilson fermions on the lattice) limit reduces to \cite{Fromm:2011qi}
\bea
\det Q_f =\prod_{\vx}\det(1+h \Tr P_\vx)^2 \det(1+\bar h \Tr P^\dagger_\vx)^2
\label{eq:hd_dirac}
\ea
where $P_\vx=\prod_{x_0=1}^{N_t} U_0(x_0,\vx)$ is the Polyakov loop (in the temporal gauge)  that is an element of  $SU(3)$ and 
\bea
h=e^{(\mu-m)/T}= (2\kappa e^{\mu a})^{N_t}, \quad \bar h=e^{(-\mu-m)/T}= (2\kappa e^{-\mu a})^{N_t}
\label{eq:fugacities}
\ea
are the quark and anti-quark fugacities with $a$, $\mu$ and $m$ being the lattice spacing, quark chemical potential and constituent quark mass. It is also useful to define  the “hopping parameter” $\kappa = e^{-ma}/2$. In the heavy-dense limit  
\bea
ma \rightarrow \infty ,\mu a \rightarrow \infty  \text{ such that } \kappa e^{\mu a} = \text{finite}\,.
\ea 
At high densities the quarks have much larger fugacity than anti-quarks, $h\gg \bar h$ and the second term in \eqnref{eq:hd_dirac} can be neglected, which simply means that
the anti-quark contribution to the effective action is negligible. We further take the lattice strong coupling limit where the gauge
field dynamics can be neglected as well. This turns out to be a fairly good approximation for $g^2_{YM} \gtrsim 1$
because as explained in Ref.  \cite{Fromm:2012eb},  the effective gauge coupling, $\lambda$, vanishes rapidly as $\lambda \sim (2N_c/g^2_{YM})^{N_t}$ for an $SU(N_c)$ gauge group 
with coupling $g^2_{YM}$. Finally, for $N_c = 3$, the Dirac determinant can be expressed in terms of $\Tr P$ as 
\bea
\det(1+hP_\vx)=(1+h\Tr P_\vx +h^2\Tr P^\dagger_\vx +h^3)
\ea
Putting everything together, to leading order in the heavy-dense limit, the partition function of the effective theory is obtained as
\bea
Z= \int[\D P] \prod_\vx(1+h\Tr P_\vx+h^2 \Tr P^\dagger_\vx+h^3)^2\,.
\label{eq:Z_free}
\ea
Apart from $h=0$  (zero density) or $h=1$ (half-filling), $P_\vx$ and $P^\dagger_\vx$  have different weights, therefore the Dirac determinant is complex and the theory has a sign problem. Furthermore the sign problem grows exponentially with volume,
\bea
\sigma=\sigma_0^V  \text{ where }\sigma_0=\frac{1 + 4 h^3 + h^6}{1 + h^2 + 2 h^3 + h^4 + h^6}
\ea
 which follows from the fact only non-vanishing group integrals are $\int [\D  P]=\int [\D   P] P^\dagger P=1$, and the path integral factorizes for each $\vx$.
  
 Beyond leading order, the dynamics of fermions can be systematically incorporated in the theory as interaction between the Polyakov loops at different spatial points and can be organized in a hopping expansion controlled by the hopping parameter $\kappa$  \cite{Langelage:2014vpa}. These interactions further simplify in the limit $N_t \gg1$, which we assume, and lead to the following  next-to-next-to-leading order  (NNLO) effective action \cite{Langelage:2014vpa}
 \begin{equation}
\label{eq:s_hopping}
\begin{split}
S_{NNLO} = & -2 \sum_{\vx} \text{Tr log}(1+h P_{\vx}) + \eta \sum_{\vx,\hat i} \text{Tr} \frac{h P_{\vx}}{1+h P_{\vx}} \text{Tr} \frac{h P_{\vxpi}}{1+h  P_{\vxpi}} 
  \\
& 
- \eta^2 \sum_{\vx,\hat i} \text{Tr} \frac{h P_{\vx}}{(1+h P_{\vx})^2} \text{Tr} \frac{h P_{\vxpi}}{(1+h P_{\vxpi})^2}
-\eta^2 \sum_{\vx,\hat{i},\hat{j}} \text{Tr} \frac{h P_{\vx}}{(1+h P_{\vx})^2} \text{Tr} \frac{h P_{\vxpi}}{(1+h P_{\vxpi})} \text{Tr} \frac{h P_{\vx+\hat{j}}}{(1+h P_{\vx+\hat{j}})}\,.
\end{split}
\end{equation}
Here $\hat i, \hat j$ denote the nearest neighbors of $\vx$ and the sums over $\hat i$ and $\hat j$ run over each direction ($\vx +\hat1, \vx -\hat 1$, etc...). The effective coupling that controls the hopping expansion is
 \bea
\eta=\frac{\kappa^2 N_t}{N_c}
 \ea
 
We have simulated the leading order as well as NNLO theory by using sign optimization.

\section{Complexification and Sign optimization}
\label{sec:sign_optimization}
In this section we summarize the idea of sign optimization \cite{Mori:2017pne,Alexandru:2018ddf,Alexandru:2018fqp,Kashiwa:2019lkv,Mori:2019tux}, and detail how we implement in heavy dense QCD.
As mentioned in the Introduction, our strategy is to complexify the path integral, without changing its value, in a way that the sign oscillations in the complexified domain are milder than the original domain. Let us denote a generic complex space whose shape depends on a set of parameters, $\vec \lambda$, as $\Ml$. Our first requirement is that complexification does not change the value of the path integral:
\bea
Z=\int_{SU(3)} [\D P] e^{-S[P]}=\int_\Ml [\D P] e^{-S[P]}\,.
\label{eq:cauchy}
\ea
This is ensured by Cauchy's theorem as long as the deformation from $SU(3)$ to $\Ml$ does not cross any singularities. The strength of the sign oscillations are captured by the average phase,
\bea
\sigma_\vl= \frac{\int_\Ml [\D P] |e^{-S[P]}|e^{-i\im S[P]}}{\int_\Ml [\D P] |e^{-S[P]}|}= \frac{\int_\Ml [\D P] e^{-S[P]}}{\int_\Ml [\D P] e^{-\re S[P]}\,.}
\ea
When the sign oscillations are strong $|\sigma_\vl|\approx 0$ due to the phase cancellations and when they are mild $|\sigma_{\vl}|\approx 1$. Notice that even though the path integral remains unchanged with the deformation, the average sign \textit{does} change, since the integrand in the denominator involves a non-holomorphic term $\re S$.  Therefore our goal is to find a set of $\vl$s which maximizes $|\sigma_{\vl}|$. To do this, we follow a gradient descent trajectory where we start from $SU(3)$ (which corresponds to $\lambda_i=0$) and update $\vl$ according to
\begin{equation}
\label{eq:grad_ascent}
\vl(\tau+1) = \vl(\tau)+ \delta \nabla_{\vl}\log |\sigma_{\vl(\tau)}|  
\end{equation}
where $\tau$ denotes the gradient ascent step. With an appropriate choice of the step size $\delta$ that is determined empirically, this procedure converges to a local maximum of $|\sigma_\vl|$.

It is clear that the complexified path integration domain, ${\cal M}\equiv \Ml^V$, must have the same dimensions as $SU(3)^V$, meaning it is a middle dimensional complex manifold embedded in $SL(3)^V$. To parameterize $\Ml$ we first express $P \in SU(3)$, with 8 angles a l\'a Bronzan \cite{Bronzan:1988wa}:
\begin{equation} \label{matrix}
P(\theta)=
\begin{pmatrix}
c_1 c_2 e^{i \theta_4} & s_1 e^{i \theta_6} & c_1 s_2 e^{i \theta_7} \\
s_2 s_3 e^{-i(\theta_7+\theta_8)}-s_1 c_2 c_3 e^{i(\theta_4+\theta_5-\theta_6)} & c_1 c_3 e^{i \theta_5} & -c_2 s_3 e^{-i(\theta_4+\theta_8)} - s_1 s_2 c_3 e^{i(\theta_5-\theta_6+\theta_7)} \\
-s_1c_2s_3 e^{i(\theta_4-\theta_6+\theta_8)}-s_2c_3 e^{-i(\theta_5+\theta_7)} & c_1s_3 e^{i\theta_8} & c_2c_3 e^{-i(\theta_4+\theta_5)}-s_1s_2s_3 e^{i(-\theta_6+\theta_7+\theta_8)} \\
\end{pmatrix}
\end{equation}
The path integral is evaluated on the domain $\theta_{i}\in [0,\pi/2]$ for $i=1,2,3$ and $\theta_i\in[0,2\pi]$ for $i=4,\dots,8$.  In this representation, the group measure is explicitly written as 
\begin{eqnarray}
dP= H(\theta) d^8\theta 
\end{eqnarray}
where $H(\theta)$ is the Haar measure, identified with the invariant measure over the $SU(3)$ manifold
\bea
H(\theta)=\sqrt{\det g}=\frac{1}{2\pi^5} s_1 c_1^3 s_2 c_2 s_3 c_3\,.
\ea
Here $g$ is the invariant $SU(3)$ metric defined as $g_{ij}=\Tr \left(P^{-1} \frac{\del P}{\del \theta_i} P^{-1} \frac{\del P}{\del \theta_j}\right)$ up to an arbitrary normalization that can be fixed by demanding $\int d^8\theta H=1$. The path integral in terms of the $\theta$ variables can be written as
\bea
Z=\int[\D p] e^{-S[P]}=\int \dth e^{-S_{eff}(\theta)} 
\ea
where
\bea
S_{eff}(\theta)=S[P(\theta)]-\sum_\vx \log H(\theta_\vx)
\label{eq:S_eff_free}
\ea
 We can now express $\Ml$ in terms of complex $\theta$s, which we denote as $\tt$. It is convenient to express the  $\tt$ in terms of its real part, which we simply call $\theta$, as
\bea
\tt_i(\theta)=\theta_i+ i f_i(\theta)
\label{eq:complexification}
\ea
where $f_i(\theta)$ are non-singular, real functions that generically depends on a set of parameters $\vl$. By construction $P(\tt(\theta))$ is  an element of a middle-dimensional manifold  $\Ml\subset SL(3)$ since it is still parameterized by eight variables. Furthermore the path integral over this manifold $\Ml$ is equal to the original path integral over $SU(3)$. This is because $\theta$ can be smoothly deformed into $\tt$ say by considering a set of intermediate surfaces parameterized by $\theta_i+i s f_i$ with $s\in[0,1]$ without crossing any singularities in the integrand. By Cauchy's theorem this deformation does not change the value of $Z$ \cite{Alexandru:2018ddf}. 

One advantage of the ansatz \eqnref{eq:complexification} is that we can express the path integral over $\Ml$ using the original real variables $\theta$
\bea
Z=\int_{\Ml} [\D P] e^{-S[P]}=\int d^{8V}\tt e^{-S_{eff}(\tt)}= \int \dth \det J  e^{-S_{eff}\left(\tt(\theta)\right)}=  \int \dth   e^{-\tSe(\theta)}
\label{eq:Z_complex}
\ea
where 
\bea
J=\frac{\partial \tt_i}{\partial \theta_j}=\delta_{ij}+i\frac{\partial f_i}{\partial \theta_j}
\label{eq:Jac}
\ea
 is the Jacobian of the transformation, and the effective action is given as
\bea
\tSe(\theta)= S_{eff}(\tt(\theta))-\sum_\vx \left[\log\det J(\theta_\vx)+\log H\left(\tt(\theta_\vx)\right)\right]
\label{eq:S_eff_complex}
\ea
We shall consider a class of complex fields generated by the so-called ``mixing" ansatz :
\begin{eqnarray} \label{eq:f_mixing}
f_i(\theta) =
\begin{cases}
0 \quad \text{for }i=1,2,3
\\
 \sum_{m,n=0}^N \lambda_{m,n}^{(i)} \cos(m \theta_4+n\theta_5) \quad \text{for }i=4,5
 \\
 \lambda^{(i)}  \quad \text{for } i=6,7,8
\end{cases}
\end{eqnarray}
This ansatz was introduced in the study of the one-dimensional QCD and it was observed to perform better than the ``diagonal" ansatz where $\im \tilde\theta_i$ is only a function of $\theta_i$, i.e. $f_i(\theta)\equiv f_i(\theta_i)$  \cite{Basar:2022cef}. The heuristic reason for the better performance is that the mixing between $\theta_i$s capture the fluctuations around complex saddle points more accurately than the diagonal ansatz. We will elaborate more on this in Section \ref{sec:thimbles} when we compare the manifold created via sign optimization to Lefschetz thimbles. 

We note that since the only degrees of freedom of the heavy-dense effective theory are Polyakov loops, $\Tr P$, we could have used the two independent eigenvalues of $\Tr P$ to parameterize the path integration domain instead of using the full $SU(3)$ parameterized by eight variables. Instead we choose to work with the full $SU(3)$, even though it is redundant to do so, in order to keep the discussion general and pave the way for more realistic cases where the degrees of freedom are the full $SU(3)$. 
That said, in this case, one would likely need to work with a more general ansatz for the complexified fields than \eqnref{eq:f_mixing}, which can easily be realized by, for instance, taking into account the Fourier coefficient of all the $\theta_i$s. A more detailed analysis of the computational performance of different choices of the anstaze is left for future work. 

Being equipped with the representation of the path integral over $\Ml$ in terms of real $\theta_i$ given in \eqnref{eq:Z_complex}, it is straightforward to simulate the theory using standard Monte-Carlo techniques by using the effective action given in \eqnref{eq:S_eff_complex}. The Markov chain on the complex field space, $\Ml$ is generated by using real $\theta$s that parameterize $\Ml$ with respect to the probability distribution $e^{-\reS}$, and the remaining phase is reweighted. In other words the expectation value of an operator is computed as 
\begin{equation}
\av{{\cal O}}=\frac1Z{\intth e^{-S_{eff}(\theta)}\,{\cal O} }= \frac1{\sigma_\vl}\frac1Z_{pq}{\intth e^{-\reS}\, {\cal O}e^{-i\imS} }=\frac1{\sigma_{\vl}}{\av{{\cal O} e^{-i\imS}  }_{\reS}}
\label{eq:lambda_reweight}
\end{equation}
where $\av{\dots}_{\reS}$ denotes the average with respect to the Boltzmann factor $e^{-\reS}$, and the phase quenched partition function and the average phase are given as 
\bea
Z_{pq}=\intth e^{-\reS},\quad \sigma_\vl=\av{e^{-i\imS}}_{\reS}\equiv \frac{\intth e^{-\tSe}}{\intth e^{-\reS}}\,.
\ea
We kept the subscript $\vl$ in the average phase to emphasize that it explicitly depends on $\vl$, even though the partition function, and therefore the expectation values of physical quantities do not. We now come back to the problem of choosing the optimal manifold within our ansatz that maximizes $\sigma_\vl$ and hence alleviates the sign problem. As mentioned earlier, the optimal choice of $\vl$ is determined via gradient ascent. We start from $SU(3)$ ($\lambda_i=0$) and update $\vl$ according to \eqnref{eq:grad_ascent}. The gradient in \eqnref{eq:grad_ascent} can be explicitly calculated as \cite{Alexandru:2018ddf}
\begin{eqnarray}
\label{eq:grad_sigma}
\gradvl \log | \sigma_{\vec{\lambda}}| &=& 
-\left\langle\intth \gradvl  e^{-\reS}\right\rangle_{\reS}= \left\langle\gradvl \re(S_{eff}-\Tr\log J) \right\rangle_{\reS} \\
&=& \left\langle \re\left(i\frac{\partial S_{eff}}{\partial \tt_i}\gradvl f_i-\Tr(J^{-1}\gradvl J)\right) \right\rangle_{\reS}
\end{eqnarray}
Therefore each gradient ascent step includes the Monte-Carlo computation of the expectation value above which, notably, can be carried out without a sign problem. 
We have simulated the theory for various parameters, both to leading order and next-to-next-to leading order using  sign optimization as outlined above. In the next section we present our results.

\section{Results}
\label{sec:results}

As explained in the previous section, the configurations on the sign optimized manifold can be sampled by using the effective action \eqnref{eq:S_eff_complex} whose argument are real $\theta_i$s, and reweighing the remaining phase, \eqnref{eq:lambda_reweight}, as usual. We simulated the theory using a Hybrid-Monte-Carlo (HMC) algorithm \cite{hmc} where we used a standard leapfrog integrator for the evolution of the Hamiltonian $h(\theta,p)=\tilde S(\theta)+p^2/2$, and used reflective boundary conditions for $\theta_{1,2,3}$ \cite{chevallier:hal-01919855}. We fixed the number of points in the Euclidean time direction to be $N_t=100$ which corresponds to the cold limit. We simulated the theory both to leading order and to next-to-next-to leading order in hopping expansion. In the former case, we fixed the hopping parameter to $\kappa=0.01$ and in the latter we varied it between $0$ and $0.05$. Note that the range of the chemical potential where the fugacity, $h$, varies between $0$ and $1$ is roughly $|1-\mu|\sim 1/N_t$ that can be seen from \eqnref{eq:fugacities}.
\begin{figure}
\center
\includegraphics[scale=0.52]{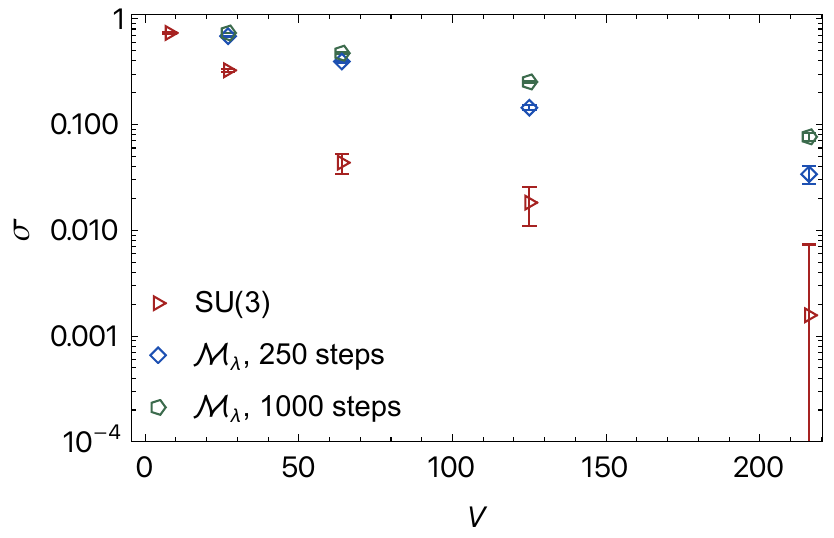}\,\,\,
\includegraphics[scale=0.5]{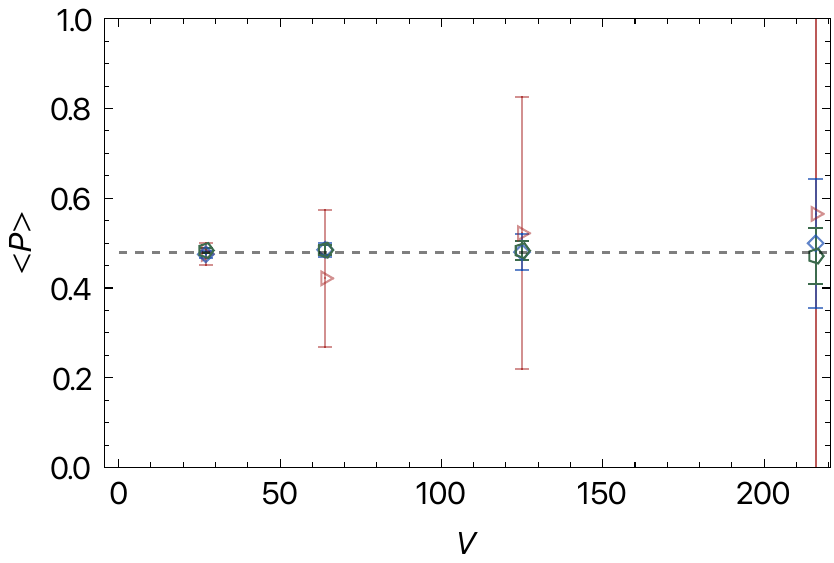}
\caption{The average sign (left) and the Polyakov loop (right) as a function of volume computed on the real plane and sign optimized manifold, ${\cal M}_\lambda$, with 250 and 1000 gradient ascent steps. The dashed gray line on the left denotes the exact value of the Polyakov loop. }
\label{fig:volume}
\end{figure}

\begin{figure}
\center
\includegraphics[scale=0.55]{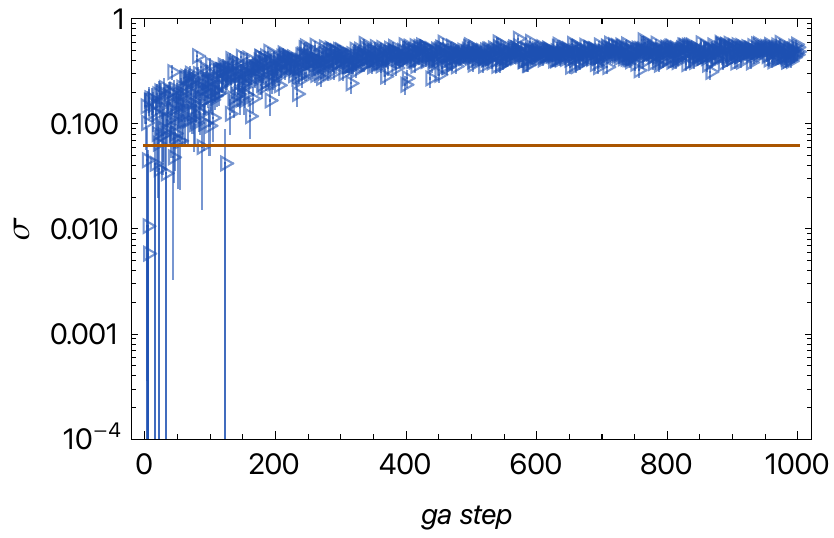}
\includegraphics[scale=0.55]{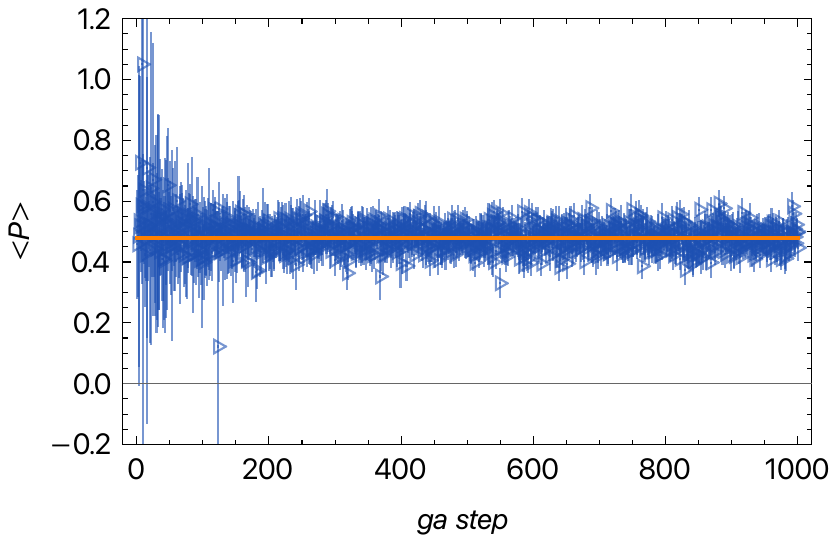}
\caption{ The average sign (left) and Polyakov loop (right) as a function of gradient ascent step for $h=0.45$ ($\kappa=0.01, \mu=0.998, N_t=100$) and $V=4^3$. The solid lines denote the exact values of $\sigma_{SU(3)}$  and $\av{P}$. }
\label{fig:signvga}
\end{figure}

In Fig. \ref{fig:volume} (left) we show the average sign as a function of volume for $\mu=0.998$, $\kappa=0.01$ and $N_t=100$ (which corresponds to fugacity $h=0.45$) to leading order in the hopping expansion. Especially for larger volumes $V=5^3$, and $6^3$ the original sign problem is bad enough where reweighting is practically not possible anymore. With sign optimization, it is lifted to values where reweighting is possible. To illustrate this, we plotted the Polyakov loop for the same values  on the right. The result after sign optimization agrees with that on the real manifold, albeit with significantly smaller error bars. The improvement in the sign problem as a function of gradient ascent is shown in Fig. \ref{fig:signvga} where we used the same parameters as above with volume  $V=4^3$. Similarly, the value of the average Polyakov loop as a function of the ascent step is shown on the right. Notice that initially the statistical uncertainty is quite substantial due to due to the severe sign problem. Gradient ascent indeed stabilizes the sign which in turn decreases the statistical uncertainty in the physical observable, $\av{P}$.

\begin{figure}
\center
\includegraphics[scale=0.42]{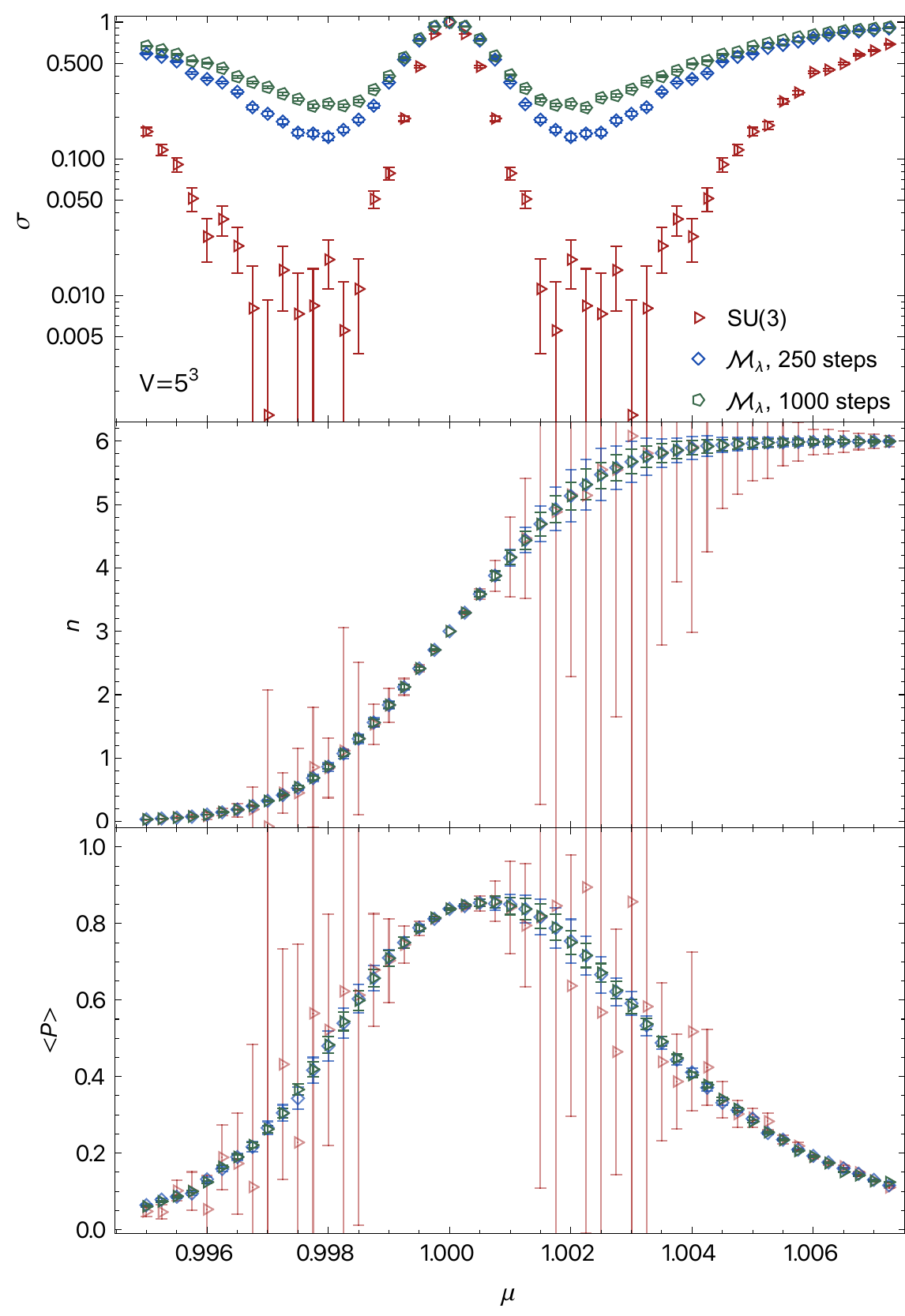}
\includegraphics[scale=0.42]{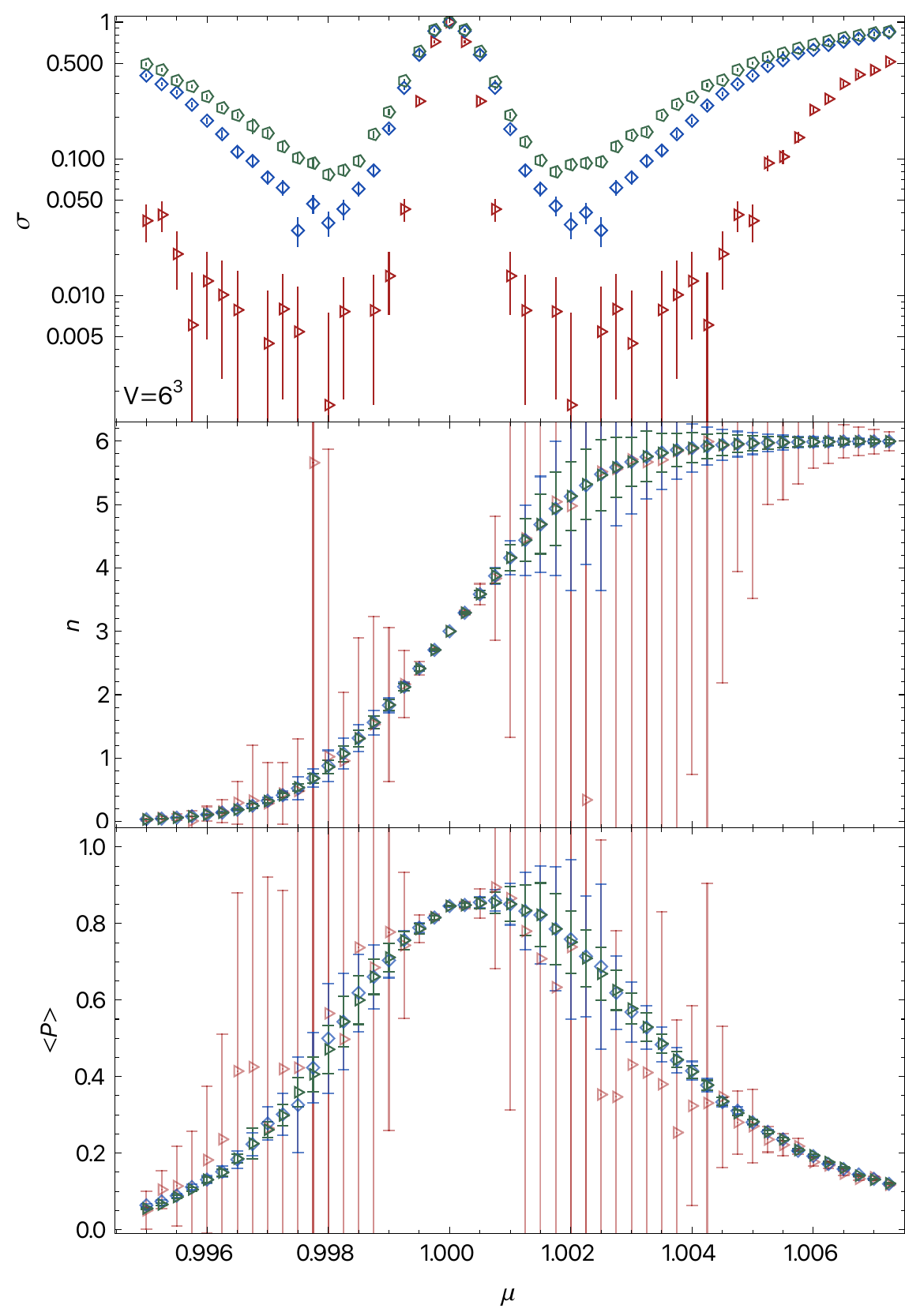}
\caption{The average sign, density and Polyakov loop as a function of chemical potential for volumes $V=5^3$ and $6^3$. }
\label{fig:free_results}
\end{figure}

The average sign, equation of state and the Polyakov loop as a function of the chemical potential for volumes $5^3$ and $6^3$ are shown in Fig \ref{fig:free_results}. For both volumes without sign optimization, the sign problem is too severe as seen from both directly from the top figures, or the error bars in the physical observables below, density and Polyakov loop. For both volumes sign optimization works as expected by stabilizing the sign and therefore significantly reducing the statistical uncertainty for all values of $\mu$. Note that the Silver Blaze phenomenon \cite{Cohen:2003kd}, where the density sharply rises when the (baryon) chemical potential reaches the baryon mass, is correctly reproduced via sign optimization, in line with previous work on heavy-dense QCD \cite{Langelage:2014vpa,Attanasio:2015Cg}. In our case this happens at $\mu=1$, where recall that that the chemical potential is measured in units of constituent quark mass.

\begin{figure}
\center
\includegraphics[scale=0.55]{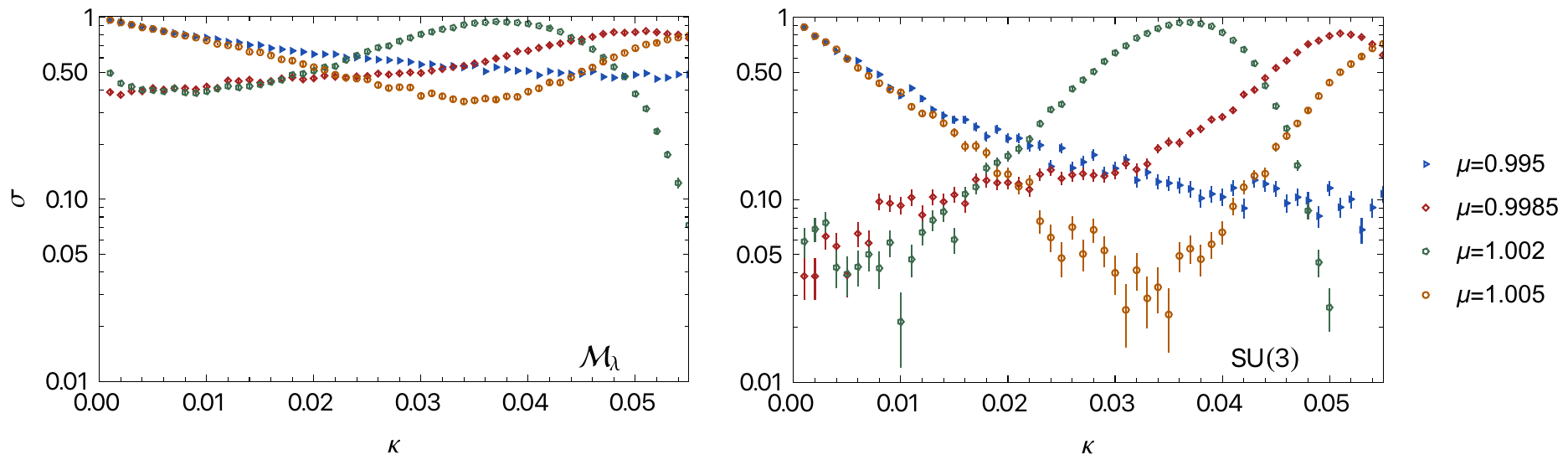}
\caption{The average sign as a function of the hopping parameter, next-to-next-to leading order in the hopping expansion with (left) and without (right) sign optimization for different values of $\mu$, $V=4^3$ and $N_t=100$.}
\label{fig:int_sign}
\end{figure}

\begin{figure}
\center
\includegraphics[scale=0.55]{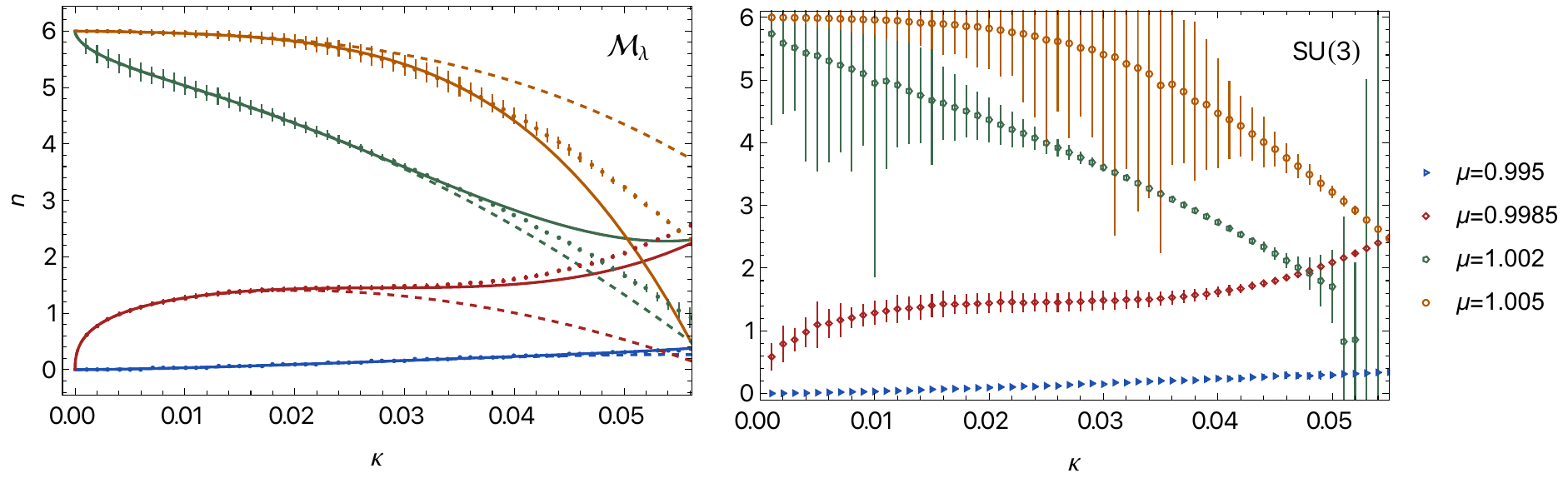}
\includegraphics[scale=0.56]{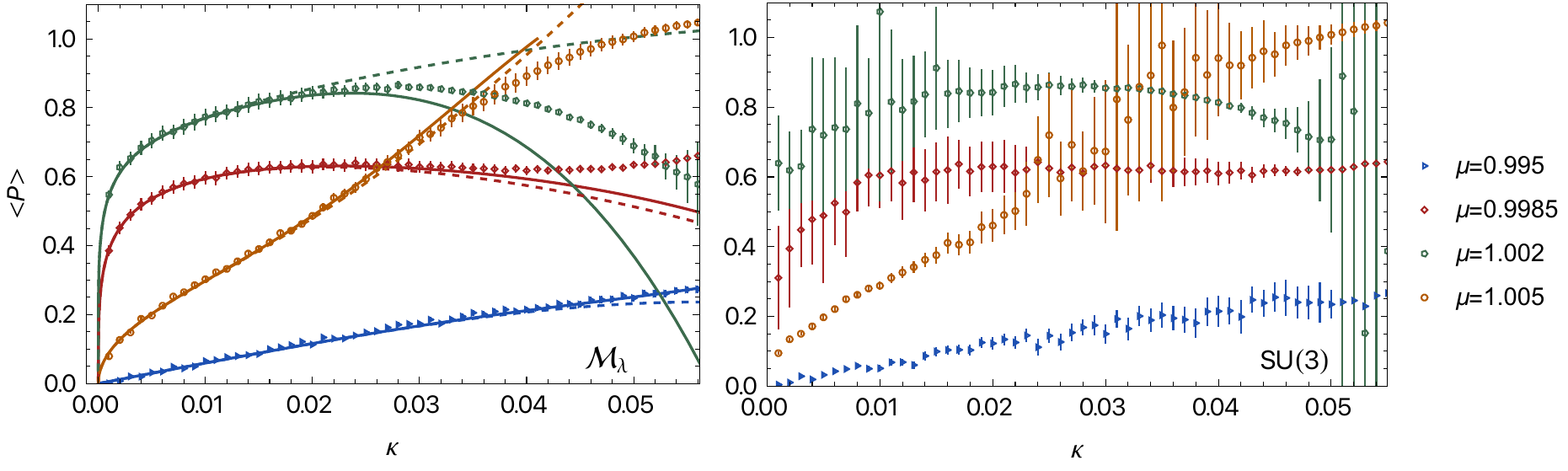}
\caption{The density and Polyakov loop as function of the hopping parameter, next-to-next-to leading order in the hopping expansion with (left) and without (right) sign optimization for different values of $\mu$, $V=4^3$ and $N_t=100$.}
\label{fig:int_n_p}
\end{figure}

Finally, in figures \ref{fig:int_sign} and \ref{fig:int_n_p} we show the average sign, density, and Polyakov loop as a function of $\kappa$, before (right) and after (left) sign optimization to next-to-next-to leading order in hopping expansion. For these simulations, we used the same form of the effective action given in \eqnref{eq:S_eff_complex} and  \eqnref{eq:S_eff_free} where in this case $S[P(\theta)]$ in \eqnref{eq:S_eff_free} is identified with the NNLO action, $S_{NNLO}[P(\theta)]$ given in \eqnref{eq:s_hopping}.
Note that for different values of the chemical potential, the sign problem becomes severe for different values of the hopping parameter $\kappa$ as seen in Fig. \ref{fig:int_sign}. As seen from the same figure, sign optimization alleviates the sign problem as expected even in the presence of the hopping terms. This is in contrast with the case where the sign problem can be solved by the worm algorithm only in to leading order (free) in hopping expansion and in parallel with Complex Langevin which also works in either case.

Finally in Fig. \ref{fig:int_n_p} we show the dependence of the density and Polyakov loop to the hopping parameter for the same values of the chemical potential as above. Similar with the free case, the statistical uncertainties that arise from the sign problem are significantly reduced by sign optimization. In order to check the validity of our results we compared them with second order perturbation theory. The dashed/solid lines show analytic results to first/second order perturbation theory. These analytic results were calculated by expanding $e^{-S_{NNLO}}$ to first (second) order in $\eta$ and evaluating the integrals over $L_\vx = \text{Tr}P_\vx$ and $L^*_\vx = \text{Tr}P_\vx^\dagger$ by the help of the relations 
\begin{eqnarray}
\text{Tr} \frac{h P_{\vec{x}}}{1+h P_{\vec{x}}} &=& \frac{h L_{\vec{x}}+2h^2 L^*_{\vec{x}}+3h^3}{1+h L_{\vec{x}}+h^2 L^*_{\vec{x}}+h^3} \\
\text{Tr} \frac{h P_{\vec{x}}}{(1+h P_{\vec{x}})^2} &=& \frac{hL_{\vec{x}}+4h^2 L^*_{\vec{x}}+9h^3}{1+h L_{\vec{x}}+h^2 L^*_{\vec{x}}+h^3}-\left( \frac{h L_{\vec{x}}+2h^2 L^*_{\vec{x}}+3h^3}{1+h L_{\vec{x}}+h^2 L^*_{\vec{x}}+h^3} \right)^2
\end{eqnarray}
 As expected the agreement between the sign optimized Monte-Carlo computations and the perturbative calculations is fairly good for small values of $\kappa$.

\section{Connection with Lefschetz thimbles}
\label{sec:thimbles}

Although being computationally very effective, by construction sign optimization does not provide much physical insight in how it alleviates the sign problem. In this section we provide a physical interpretation of the results we presented by  making a connection between the sign optimized manifold, $\Ml$, and the Lefschetz thimbles associated with the theory. 
In order to construct the Lefschetz thimbles associated with the path integral, \eqnref{eq:Z_free}, to compare and contrast with $\Ml$, we first express the path integral in terms of the eigenvalues of the Polyakov loop, $\alpha_1$, $\alpha_2$. We choose to work with $\alpha$s instead of $\theta$s in order to visualize the connection with the thimbles easily.  The Polyakov loop in the diagonal form is given as $P={\rm diag}\{e^{i\alpha_1}, e^{i\alpha_2}, e^{-i(\alpha_1+\alpha_2)}\}$. The group measure  for the $\alpha$ variables is a Vandermonde determinant:
\bea
V(\vec \alpha)=\prod_{i<j}|e^{i\alpha_i}-e^{i\alpha_j}|=\sin^2\left(\frac{\alpha_1-\alpha_2}{2}\right) \sin^2\left(\frac{\alpha_1+2\alpha_2}{2}\right) \sin^2\left(\frac{2\alpha_1+\alpha_2}{2}\right) 
\ea
such that the partition function is expressed as 
\bea
Z=\int  \prod_{\vx}\left[ d\alpha_{1,\vx}d\alpha_{2,\vx} V(\alpha_{\vx})\right] e^{-S[P(\vec\alpha_\vx)]}=\int d^{2V}\alpha \,\,e^{-S_e[\vec\alpha_\vx]}\,.
\ea
Here the effective action is defined as 
\bea
S_{e}[\vec\alpha_\vx]=\sum_{\vx}\left[-2\log \left(1+ h\Tr P(\vec\alpha_\vx)  +h^2 P^\dagger(\vec\alpha_\vx)+h^3\right)+\log V(\vec \alpha_\vx) \right]
\label{eq:Se}
\ea
and the Polyakov loop takes the form $\Tr P(\vec\alpha)=e^{i\alpha_1}+ e^{i\alpha_2}+ e^{-i(\alpha_1+\alpha_2)}$. 

The Lefschetz thimble decomposition of $Z$ can be constructed by evaluating the holomorphic gradient flow 
\bea
\frac{d \alpha_{i,\vx}}{d\tau}=\overline{\frac{\partial S[\vec \alpha_\vx]}{\partial \alpha_{i,\vx}}}
\label{eq:flow}
\ea
starting from $\alpha_{i,\vx}\in [0,2\pi]$ and taking the formal limit $\tau\rightarrow \infty$. Finite values of $\tau$ are associated with a family of complex manifolds that interpolate between the original domain, $[0,2\pi]^{2V}$, and the thimble decomposition.  Following \cite{Alexandru:2015sua}, we shall call these manifolds "generalized thimbles". 

Since there are no interactions between different spatial points in the leading order in hopping expansion given in \eqnref{eq:Z_free}, the flow equation,  \eqnref{eq:flow} can be solved for each spatial point, $\vx$, independently. Therefore each generalized thimble is $V$ dimensional direct product of a two dimensional complex manifold. We solve the flow equation to generate this two dimensional manifold for different flow times. 

In order to visualize the generalized thimble, we then take a projection on the subspace defined by $\alpha_1+\alpha_2^*=0$. Notably, this quantity is ``conserved" under the flow, namely 
\bea
\frac{d }{d\tau}(\alpha_1+\alpha_2^*)=0\,.
\label{eq:conservation}
\ea
This can be seen by observing that the action, $S_e$, given in Eq. \eqref{eq:Se} is a real function of $x$ and $y$ where $\alpha_1=-\alpha_2^*=x+iy$. The conservation equation,  \eqnref{eq:conservation},  then follows directly from the flow equation, \eqnref{eq:flow}. 
\begin{figure}
\center
\includegraphics[scale=0.6]{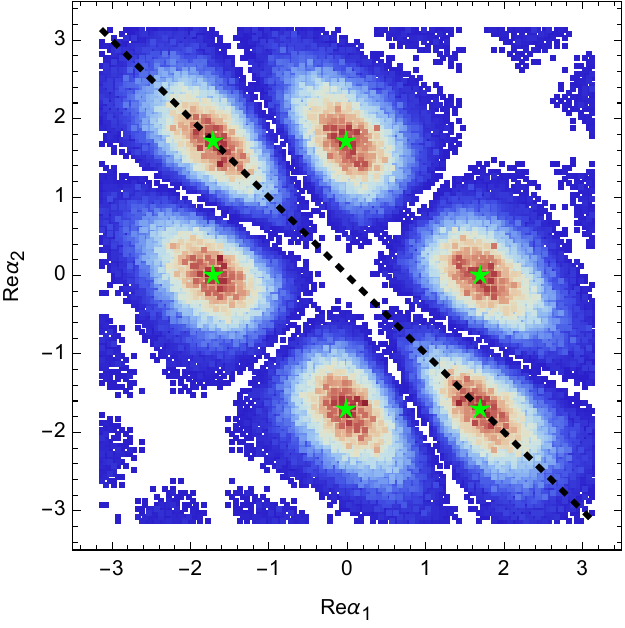}
\includegraphics[scale=0.62]{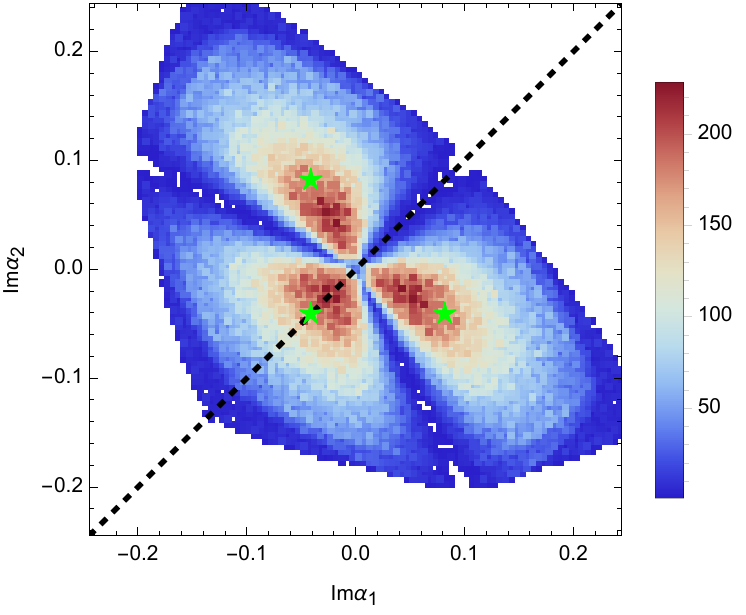}
\caption{The histogram of the eigenvalues of the Polyakov loop averaged over the volume. The green stars denote the saddle points and the dashed line represents the subspace defined by $\alpha_1+\alpha_2^*=0$ which we plot the Lefschetz thimble on.}
\label{fig:thimble-hist}
\end{figure}
We now compare the sign optimized manifold and the generalized thimbles on the projection on the subspace $\alpha_1+\alpha_2^*=0$ denoted by the dashed black line in Fig. \ref{fig:thimble-hist}.  Projections of the generalized thimbles as well as the sign optimized manifold on this subspace are one-dimensional, which we show in Fig. \ref{fig:thimble-vs-opt}. 

\begin{figure}
\center
\includegraphics[scale=0.6]{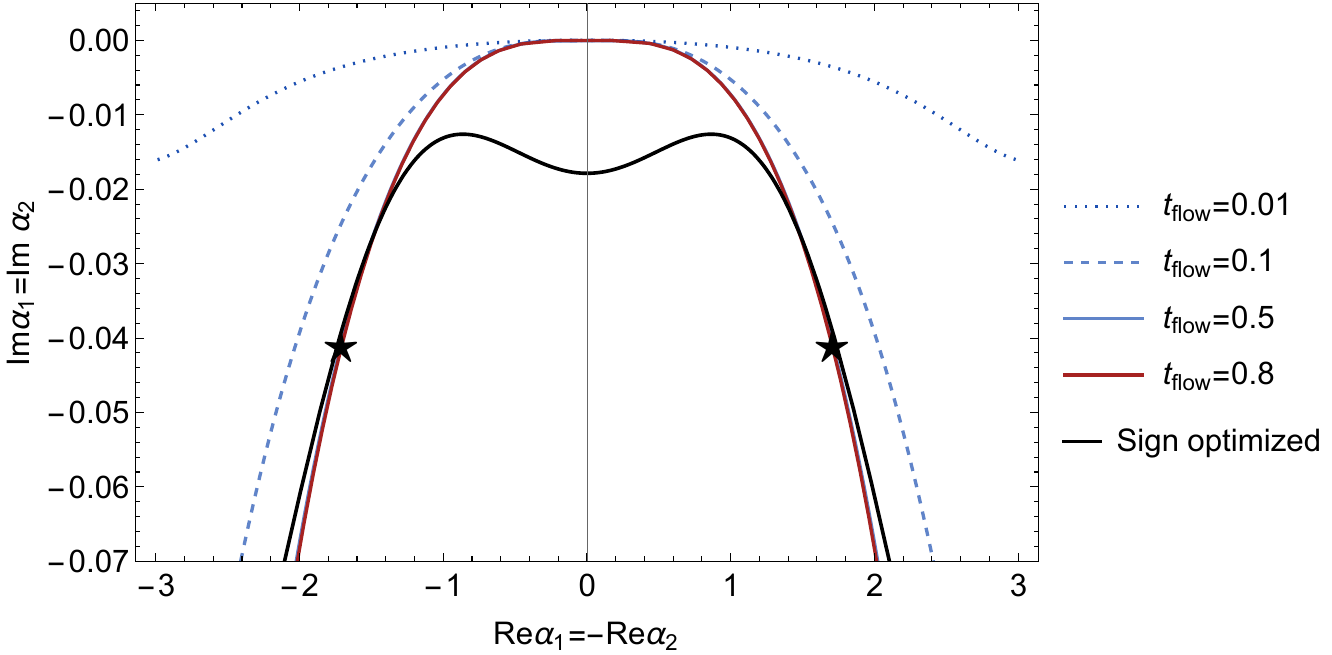}
\caption{The comparison between the surfaces generated by holomorphic gradient flow, approaching the Lefschetz thimble at large flow times, and the sign optimized manifold, plotted on the subspace defined by $\alpha_1+\alpha_2^*=0$. The stars denote the saddle points that lie on this subspace (see Fig \ref{fig:thimble-hist})}
\label{fig:thimble-vs-opt}
\end{figure}

Firstly, it can be seen that the generalized thimbles converge into the the thimble decomposition with increasing flow time as expected. Secondly, there are multiple complex saddle points, hence multiple thimbles, which contribute to the thimble decomposition. There are two saddle points on the subspace that we focus on (those that intersect the dashed line in  Fig. \ref{fig:thimble-hist}), but the other saddle points contribute equally as well since $\re\, S_e$ (hence the path integral weight) on each saddle point is equal. Thirdly, the sign optimized manifold approximately reconstructs the thimble, especially around the saddle points where most of the contribution to the path integral comes from.

\begin{figure}[ht]
  \centering  
  \includegraphics[scale=.58]{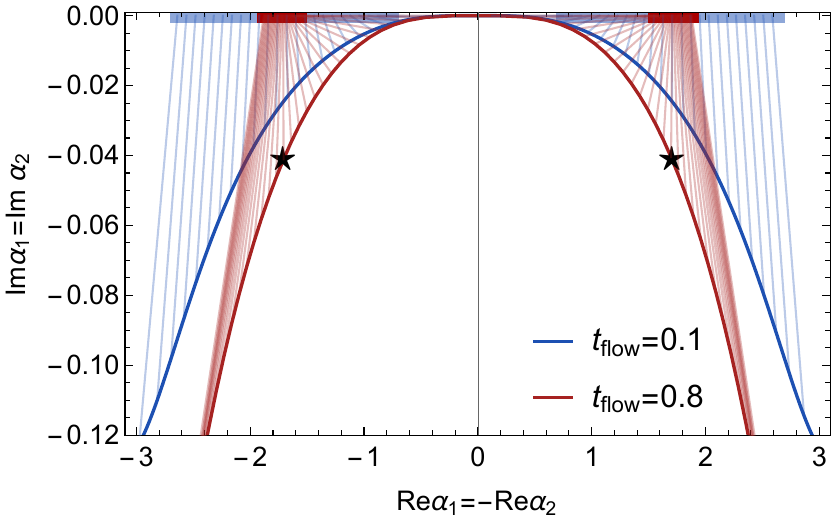}
   \includegraphics[scale=.58]{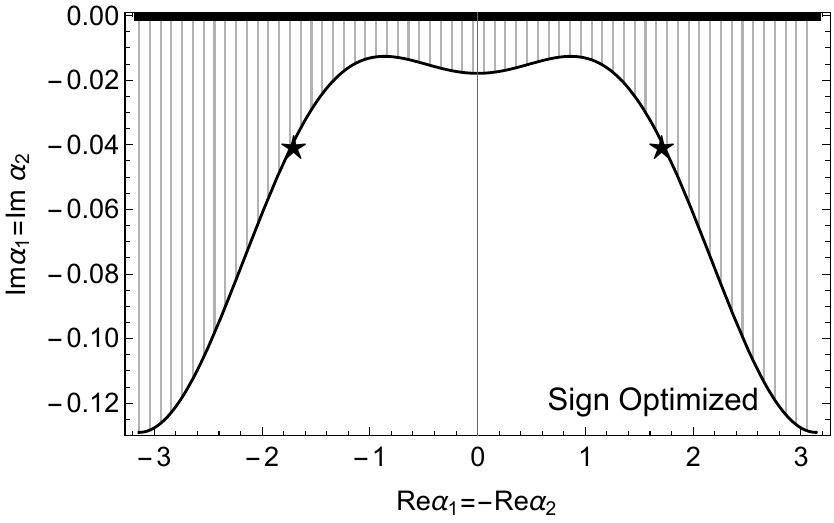}
 \caption{The parameterization of the generalized thimbles (left) and the sign optimized manifold (right) in terms of the real fields (bars). For thimbles, with increasing flow time the relevant region in the real filed space that is sampled shrinks into small, disconnected regions leading to multimodal distribution in contrast to sign optimization.}
  \label{fig:regions}
\end{figure}

 We would like to point out that in general, sampling multi-model distributions like this could be present challenges as the Markov chain could get stuck in a local minimum. The generalized thimble approach is susceptible to this phenomenon as the relevant support of each thimble in the sampled field space, $\re\,\alpha$, shrinks with increasing flow time as pictured in Fig. \ref{fig:regions}. As a result transitioning from one such region to another becomes more difficult. The sign optimization, however, does not. This follows directly from the form of the parameterization given in Eq. \eqref{eq:f_mixing} as $\im\,\alpha_i$ is expressed as a function of $\re\,\alpha_i$, so that the size of the relevant region in the sampled field space remains the same \cite{Alexandru:2020wrj}.  Finally, the Jacobian associated with the sign optimization ansatz, Eq. \eqref{eq:Jac}, is fairly simple and can be analytically calculated, as opposed to the Jacobian associated with holomorphic gradient flow which has to be numerically evaluated, introducing extra computational cost.   

\section{Conclusions}
\label{sec:conclusions}

We simulated the heavy-dense effective theory of QCD at finite density by using sign optimization to alleviate the sign problem. The optimal complex domain for the path integral is chosen within a family of anstaze via a gradient ascent algorithm. Building up on our previous observation from one-dimensional QCD, we used the ``mixing" ansatz for the complexified manifold which is constructed to capture the fluctuations of fields around the complex saddle points. We have simulated the theory for different values of the hopping parameter, volume and chemical potential both to leading order and next-to-next-to leading order in the hopping expansion. Notably the sign problem in the latter case cannot be solved via worm algorithm as opposed to the former case \cite{DelgadoMercado:2012gta}.  We observed that the sign optimization method can handle both cases with comparable amount of computational resources. We also demonstrated that sign optimization works successfully for volumes $4^3$, $5^3$, and $6^3$ where the sign problem is too severe to overcome with usual reweighing.

Even though sign optimization is demonstrated to be an efficient way to alleviate the sign problem, it is not transparent how it does it; i.e. it does not provide much of physical insight. This is, of course, a generic property of optimization algorithms. We showed that sign optimization in fact in a way reconstructs the Lefschetz thimbles (a multi-dimensional generalization of stationary phase contours) around the complex saddle points of the theory. Similar phenomena has been observed before \cite{Alexandru:2018fqp}. In this paper we explicitly compared the Lefschetz thimble and the sign optimized manifold and showed that they indeed overlap especially in the vicinity of the saddle points. This comparison further solidifies the observation of the similarity between the Lefschetz thimbles emanating from the complex saddles of the theory and the sign optimized manifold made in one-dimensional QCD \cite{Basar:2022cef}.  As a result, one might view sign optimization is a method to construct an approximation to the Lefschetz thimbles. However, the there are advantages of  using sign optimization instead of directly sampling the (generalized) Lefschetz thimbles. Firstly, the parameterization of the sign optimized manifold does not require a numerical computation of the Jacobian as opposed to thimbles as it can be performed analytically, reducing the computational cost. Secondly sampling the sign optimized manifold does not lead to severe multi-modal distributions, which are difficult to sample as opposed to the thimbles.

We performed sign optimization within a family of complex manifolds which are parameterized by a fairly generic ansatz that can in principle be applied to an arbitrary theory whose degrees of freedom take value in $SU(3)$. This generality comes with a cost, however. Finding a \textit{generic} solution to the sign problem is an NP hard problem \cite{Troyer:2004ge}, therefore any generic ansatz, such as we used, will still lead to an exponentially hard problem. Of course, even reducing the exponent lead be practically useful results as we demonstrated here.  At the same time, by using certain properties of the underlying theory such as symmetries or perhaps some other knowledge such as complex saddle points, one could consider  a more \textit{specific} ansatz tailored for the specific theory in mind. In this case, there is no a-priori reason for the sign problem to be exponentially hard. Here, we took the first step in relating the sign optimized manifolds to the complex saddle points and the fluctuations around them. The natural next step is to improve the complexification ansatz by using some analytical information based on these observations, which is left for future work.

\acknowledgments
GB is supported by the National Science Foundation CAREER Award PHY-2143149.

\bibliographystyle{utphys}
\bibliography{refs}

\providecommand{\href}[2]{#2}\begingroup\raggedright\begin{thebibliography}{10}

\bibitem{HotQCD:2014kol}
{\bfseries HotQCD} Collaboration, A.~Bazavov {\em et~al.}, ``{Equation of state
  in ( 2+1 )-flavor QCD},''
  \href{http://dx.doi.org/10.1103/PhysRevD.90.094503}{{\em Phys. Rev. D}
  {\bfseries 90} (2014) 094503},
  \href{http://arxiv.org/abs/1407.6387}{{\ttfamily arXiv:1407.6387 [hep-lat]}}.

\bibitem{BORSANYI201499}
S.~Bors{\'a}nyi, Z.~Fodor, C.~Hoelbling, S.~D. Katz, S.~Krieg, and K.~K.
  Szab{\'o}, ``Full result for the qcd equation of state with 2+1 flavors,''
  \href{http://dx.doi.org/https://doi.org/10.1016/j.physletb.2014.01.007}{{\em
  Physics Letters B} {\bfseries 730} (2014) 99--104}.
  \url{https://www.sciencedirect.com/science/article/pii/S0370269314000197}.

\bibitem{deForcrandreview}
P.~de~Forcrand, ``Simulating qcd at finite density,'' 2010.

\bibitem{Philipsen:2007aa}
O.~Philipsen, ``Lattice qcd at finite temperature and density,''
  \href{http://dx.doi.org/10.1140/epjst/e2007-00376-3}{{\em The European
  Physical Journal Special Topics} {\bfseries 152} no.~1, (2007) 29--60}.
  \url{https://doi.org/10.1140/epjst/e2007-00376-3}.

\bibitem{KARSCH200014}
F.~Karsch, ``Lattice qcd at finite temperature and density,''
  \href{http://dx.doi.org/https://doi.org/10.1016/S0920-5632(00)91591-3}{{\em
  Nuclear Physics B - Proceedings Supplements} {\bfseries 83-84} (2000) 14 --
  23}.
  \url{http://www.sciencedirect.com/science/article/pii/S0920563200915913}.
  Proceedings of the XVIIth International Symposium on Lattice Field Theory.

\bibitem{Aarts_2016}
G.~Aarts, ``Introductory lectures on lattice qcd at nonzero baryon number,''
  \href{http://dx.doi.org/10.1088/1742-6596/706/2/022004}{{\em Journal of
  Physics: Conference Series} {\bfseries 706} (Apr, 2016) 022004}.
  \url{http://dx.doi.org/10.1088/1742-6596/706/2/022004}.

\bibitem{pham}
F.~Pham, ``{Vanishing homologies and the n variable saddlepoint method},'' {\em
  Methods Appl. Anal. 1} {\bfseries Part 2} (1983) .

\bibitem{Lefschetz:1975ta}
S.~Lefschetz, {\em {Applications of Algebraic Topology. Graphs and Networks.
  the Picard-Lefschetz Theory and Feynman Integrals}}.
\newblock 1, 1975.

\bibitem{Cristoforetti:2012su}
{\bfseries AuroraScience} Collaboration, M.~Cristoforetti, F.~Di~Renzo, and
  L.~Scorzato, ``{New approach to the sign problem in quantum field theories:
  High density QCD on a Lefschetz thimble},''
  \href{http://dx.doi.org/10.1103/PhysRevD.86.074506}{{\em Phys. Rev. D}
  {\bfseries 86} (2012) 074506},
  \href{http://arxiv.org/abs/1205.3996}{{\ttfamily arXiv:1205.3996 [hep-lat]}}.

\bibitem{Cristoforetti:2012uv}
M.~Cristoforetti, L.~Scorzato, and F.~Di~Renzo, ``{The sign problem and the
  Lefschetz thimble},'' \href{http://arxiv.org/abs/1210.8026}{{\ttfamily
  arXiv:1210.8026 [hep-lat]}}.
[J. Phys. Conf. Ser.432,012025(2013)].

\bibitem{Fujii:2013sra}
H.~Fujii, D.~Honda, M.~Kato, Y.~Kikukawa, S.~Komatsu, and T.~Sano, ``{Hybrid
  Monte Carlo on Lefschetz thimbles - A study of the residual sign problem},''
  \href{http://dx.doi.org/10.1007/JHEP10(2013)147}{{\em JHEP} {\bfseries 10}
  (2013) 147}, \href{http://arxiv.org/abs/1309.4371}{{\ttfamily arXiv:1309.4371
  [hep-lat]}}.

\bibitem{Alexandru:2020wrj}
A.~Alexandru, G.~Basar, P.~F. Bedaque, and N.~C. Warrington, ``{Complex paths
  around the sign problem},''
  \href{http://dx.doi.org/10.1103/RevModPhys.94.015006}{{\em Rev. Mod. Phys.}
  {\bfseries 94} no.~1, (2022) 015006},
  \href{http://arxiv.org/abs/2007.05436}{{\ttfamily arXiv:2007.05436
  [hep-lat]}}.

\bibitem{Mori:2017pne}
Y.~Mori, K.~Kashiwa, and A.~Ohnishi, ``{Toward solving the sign problem with
  path optimization method},''
  \href{http://dx.doi.org/10.1103/PhysRevD.96.111501}{{\em Phys. Rev. D}
  {\bfseries 96} no.~11, (2017) 111501},
  \href{http://arxiv.org/abs/1705.05605}{{\ttfamily arXiv:1705.05605
  [hep-lat]}}.

\bibitem{Alexandru:2018fqp}
A.~Alexandru, P.~F. Bedaque, H.~Lamm, and S.~Lawrence, ``{Finite-Density Monte
  Carlo Calculations on Sign-Optimized Manifolds},''
  \href{http://dx.doi.org/10.1103/PhysRevD.97.094510}{{\em Phys. Rev. D}
  {\bfseries 97} no.~9, (2018) 094510},
  \href{http://arxiv.org/abs/1804.00697}{{\ttfamily arXiv:1804.00697
  [hep-lat]}}.

\bibitem{Kashiwa:2019lkv}
K.~Kashiwa, Y.~Mori, and A.~Ohnishi, ``{Application of the path optimization
  method to the sign problem in an effective model of QCD with a repulsive
  vector-type interaction},''
  \href{http://dx.doi.org/10.1103/PhysRevD.99.114005}{{\em Phys. Rev.}
  {\bfseries D99} no.~11, (2019) 114005},
\href{http://arxiv.org/abs/1903.03679}{{\ttfamily arXiv:1903.03679 [hep-lat]}}.

\bibitem{Detmold:2021ulb}
W.~Detmold, G.~Kanwar, H.~Lamm, M.~L. Wagman, and N.~C. Warrington, ``{Path
  integral contour deformations for observables in $SU(N)$ gauge theory},''
  \href{http://dx.doi.org/10.1103/PhysRevD.103.094517}{{\em Phys. Rev. D}
  {\bfseries 103} no.~9, (2021) 094517},
  \href{http://arxiv.org/abs/2101.12668}{{\ttfamily arXiv:2101.12668
  [hep-lat]}}.

\bibitem{Zambello:2018ibq}
K.~Zambello and F.~Di~Renzo, ``{Towards Lefschetz thimbles regularization of
  heavy-dense QCD},'' \href{http://dx.doi.org/10.22323/1.334.0148}{{\em PoS}
  {\bfseries LATTICE2018} (2018) 148},
  \href{http://arxiv.org/abs/1811.03605}{{\ttfamily arXiv:1811.03605
  [hep-lat]}}.

\bibitem{Attanasio:2015Cg}
F.~Attanasio, G.~Aarts, B.~Jaeger, E.~Seiler, D.~Sexty, and I.-O. Stamatescu,
  ``{Exploring the phase diagram of QCD with complex Langevin simulations},''
  \href{http://dx.doi.org/10.22323/1.214.0200}{{\em PoS} {\bfseries
  LATTICE2014} (2015) 200}.

\bibitem{Aarts:2016qrv}
G.~Aarts, F.~Attanasio, B.~J\"ager, and D.~Sexty, ``{The QCD phase diagram in
  the limit of heavy quarks using complex Langevin dynamics},''
  \href{http://dx.doi.org/10.1007/JHEP09(2016)087}{{\em JHEP} {\bfseries 09}
  (2016) 087}, \href{http://arxiv.org/abs/1606.05561}{{\ttfamily
  arXiv:1606.05561 [hep-lat]}}.

\bibitem{Basar:2022cef}
G.~Basar and J.~Marincel, ``{Sign optimization and complex saddle points in
  one-dimensional QCD},''
  \href{http://dx.doi.org/10.1103/PhysRevD.106.L091503}{{\em Phys. Rev. D}
  {\bfseries 106} no.~9, (2022) L091503},
  \href{http://arxiv.org/abs/2208.02072}{{\ttfamily arXiv:2208.02072
  [hep-lat]}}.

\bibitem{Fromm:2011qi}
M.~Fromm, J.~Langelage, S.~Lottini, and O.~Philipsen, ``{The QCD deconfinement
  transition for heavy quarks and all baryon chemical potentials},''
  \href{http://dx.doi.org/10.1007/JHEP01(2012)042}{{\em JHEP} {\bfseries 01}
  (2012) 042}, \href{http://arxiv.org/abs/1111.4953}{{\ttfamily arXiv:1111.4953
  [hep-lat]}}.

\bibitem{Fromm:2012eb}
M.~Fromm, J.~Langelage, S.~Lottini, M.~Neuman, and O.~Philipsen, ``{Onset
  Transition to Cold Nuclear Matter from Lattice QCD with Heavy Quarks},''
  \href{http://dx.doi.org/10.1103/PhysRevLett.110.122001}{{\em Phys. Rev.
  Lett.} {\bfseries 110} no.~12, (2013) 122001},
  \href{http://arxiv.org/abs/1207.3005}{{\ttfamily arXiv:1207.3005 [hep-lat]}}.

\bibitem{Langelage:2014vpa}
J.~Langelage, M.~Neuman, and O.~Philipsen, ``{Heavy dense QCD and nuclear
  matter from an effective lattice theory},''
  \href{http://dx.doi.org/10.1007/JHEP09(2014)131}{{\em JHEP} {\bfseries 09}
  (2014) 131}, \href{http://arxiv.org/abs/1403.4162}{{\ttfamily arXiv:1403.4162
  [hep-lat]}}.

\bibitem{Alexandru:2018ddf}
A.~Alexandru, P.~F. Bedaque, H.~Lamm, S.~Lawrence, and N.~C. Warrington,
  ``{Fermions at Finite Density in 2+1 Dimensions with Sign-Optimized
  Manifolds},'' \href{http://dx.doi.org/10.1103/PhysRevLett.121.191602}{{\em
  Phys. Rev. Lett.} {\bfseries 121} no.~19, (2018) 191602},
  \href{http://arxiv.org/abs/1808.09799}{{\ttfamily arXiv:1808.09799
  [hep-lat]}}.

\bibitem{Mori:2019tux}
Y.~Mori, K.~Kashiwa, and A.~Ohnishi, ``{Path optimization in $0+1$D QCD at
  finite density},'' \href{http://dx.doi.org/10.1093/ptep/ptz111}{{\em PTEP}
  {\bfseries 2019} no.~11, (2019) 113B01},
  \href{http://arxiv.org/abs/1904.11140}{{\ttfamily arXiv:1904.11140
  [hep-lat]}}.

\bibitem{Bronzan:1988wa}
J.~B. Bronzan, ``{Parametrization of SU(3)},''
  \href{http://dx.doi.org/10.1103/PhysRevD.38.1994}{{\em Phys. Rev. D}
  {\bfseries 38} (1988) 1994}.

\bibitem{hmc}
S.~Brooks, A.~Gelman, G.~Jones, and X.-L. Meng, eds.,
  \href{http://dx.doi.org/10.1201/b10905}{{\em Handbook of Markov Chain Monte
  Carlo}}.
\newblock Chapman and Hall/{CRC}, May, 2011.
\newblock \url{https://doi.org/10.1201%2Fb10905}.

\bibitem{chevallier:hal-01919855}
A.~Chevallier, S.~Pion, and F.~Cazals, ``{Hamiltonian Monte Carlo with boundary
  reflections, and application to polytope volume calculations},'' Research
  Report RR-9222, {INRIA Sophia Antipolis, France}, Nov., 2018.
\newblock \url{https://hal.science/hal-01919855}.

\bibitem{Cohen:2003kd}
T.~D. Cohen, ``{Functional integrals for QCD at nonzero chemical potential and
  zero density},'' \href{http://dx.doi.org/10.1103/PhysRevLett.91.222001}{{\em
  Phys. Rev. Lett.} {\bfseries 91} (2003) 222001},
\href{http://arxiv.org/abs/hep-ph/0307089}{{\ttfamily arXiv:hep-ph/0307089
  [hep-ph]}}.

\bibitem{Alexandru:2015sua}
A.~Alexandru, G.~Basar, P.~F. Bedaque, G.~W. Ridgway, and N.~C. Warrington,
  ``{Sign problem and Monte Carlo calculations beyond Lefschetz thimbles},''
  \href{http://dx.doi.org/10.1007/JHEP05(2016)053}{{\em JHEP} {\bfseries 05}
  (2016) 053},
\href{http://arxiv.org/abs/1512.08764}{{\ttfamily arXiv:1512.08764 [hep-lat]}}.

\bibitem{DelgadoMercado:2012gta}
Y.~Delgado~Mercado and C.~Gattringer, ``{Monte Carlo simulation of the SU(3)
  spin model with chemical potential in a flux representation},''
  \href{http://dx.doi.org/10.1016/j.nuclphysb.2012.05.009}{{\em Nucl. Phys. B}
  {\bfseries 862} (2012) 737--750},
  \href{http://arxiv.org/abs/1204.6074}{{\ttfamily arXiv:1204.6074 [hep-lat]}}.

\bibitem{Troyer:2004ge}
M.~Troyer and U.-J. Wiese, ``{Computational complexity and fundamental
  limitations to fermionic quantum Monte Carlo simulations},''
  \href{http://dx.doi.org/10.1103/PhysRevLett.94.170201}{{\em Phys. Rev. Lett.}
  {\bfseries 94} (2005) 170201},
\href{http://arxiv.org/abs/cond-mat/0408370}{{\ttfamily arXiv:cond-mat/0408370
  [cond-mat]}}.

\end{thebibliography}\endgroup

\end{document}